\documentclass[twocolumn,twoside,slac_two]{revtex4}
\usepackage{epsfig,graphicx}
\usepackage{epstopdf}
\usepackage{fancyhdr}
\begin{document}

\newcommand {\be}{\begin{equation}}
\newcommand {\ee}{\end{equation}}
\newcommand {\ba}{\begin{eqnarray}}
\newcommand {\ea}{\end{eqnarray}}

\newcommand{\hif}{\mathchar`-}
\newcommand{\notD}{\ /\hspace{-2.5mm}D}
\newcommand{\noteq}{\ /\hspace{-3.5mm}=}
\newcommand{\tr}{{\rm tr}}
\newcommand{\Tr}{{\rm Tr}}
\newcommand{\diag}{{\rm diag}}
\newcommand{\ev}{{\rm eV}}
\newcommand{\mev}{{\rm MeV}}
\newcommand{\gev}{{\rm GeV}}
\newcommand{\tev}{{\rm TeV}}

\newcommand{\Lmd}{\Lambda}
\newcommand{\lmd}{\lambda}
\newcommand{\TeV}{\mbox{TeV}}
\newcommand{\GeV}{\mbox{GeV}}
\newcommand{\MeV}{\mbox{MeV}}
\newcommand{\SM}{\mathrm{SM}}
\newcommand{\simless}{\hspace{0.3em}\raisebox{0.4ex}{$<$}\hspace{-0.75em}\raisebox{-.7ex}{$\sim$}\hspace{0.3em}}
\newcommand{\del}{\partial}
\newcommand{\nn}{\nonumber}
\newcommand{\nin}{\noindent}
\newcommand{\tz}{{T^0}}
\newcommand{\tp}{{T^+}}
\newcommand{\tpp}{{T^{++}}}
\newcommand{\tzr}{{T_r^0}}
\newcommand{\tzi}{{T_i^0}}

\renewcommand{\thesection}{\arabic{section}}
\renewcommand{\thesubsection}{\arabic{section}.\arabic{subsection}}
\renewcommand{\thefigure}{\arabic{figure}}

\title{\bf Footprint of Triplet Scalar Dark Matter in Direct, Indirect Search and Invisible Higgs Decay }

\author{${\rm Seyed~ Yaser~ Ayazi}$$^1$ and ${\rm S.~Mahdi~ Firouzabadi}$$^2$}
\affiliation{\small $^1$$School~ of~ Particles~ and~ Accelerators,~ Institute~ for~ Research~ in~
Fundamental$ $ Sciences~(IPM), P.O.~Box~ 19395-5531, Tehran, Iran$}
\affiliation{\small $^2$$Department~ of~ Physics,~ Shahid~ Beheshti~ University,$ $G.~ C.,~ Evin, ~Tehran~ 19839,~ Iran$}

\begin{abstract}
In this talk, we will review Inert Triplet Model (ITM) which provide candidate for dark matter (DM) particles. Then we study possible decays of Higgs boson to DM  candidate and apply current experimental data for invisible Higgs decay to constrain parameter space of ITM. We also consider indirect search for DM and use  FermiLAT data to put constraints on parameter space. Ultimately we compare  this limit with constraints provided by LUX experiment for low mass DM and invisible Higgs decay.
\end{abstract}

\maketitle

\thispagestyle{fancy}

\section{Introduction}
There are strong evidences for non-baryonic DM which according to Planck satellite\cite{PLANCK} constitute more than 0.26 of energy density in the universe. WIMP's as a relic remnants of early universe are the most plausible candidates for DM.
Since the standard model (SM) cannot explain DM evidences,  there is a strong motivation to extend SM in a way to provide suitable DM candidate.
Singlet scalar or fermion fields are preferred  as simple candidates for DM.  It is shown that allowed regions of parameters space for these models are strictly limited by WMAP data \cite{SDM},\cite{Kim:2006af}. One of the simplest models for a scalar dark matter is ITM. In this model, a scalar $SU(2)_L$ triplet is odd under $Z_2$ symmetry so that they can directly couple to the SM particles and the neutral components of the triplets play role of DM.

After a few decades of expectations, the LHC has found a SM like Higgs particle with a mass of $125~\rm GeV$. Since the Higgs boson can participate in DM-nucleon scattering and DM annihilation, current analysis of the LHC data  and measurements of its decay rates would set limit on any beyond SM that provides a DM candidate.

In this talk, we extend SM by a $SU(2)_L$ triplet scalar with hypercharge $Y=0,2$. The lightest component of triplet field is neutral and provides suitable candidate for DM\cite{Ayazi:2014tha}. Then, we review allowed parameters space of ITM by PLANCK data and invisible higgs decay measurement, direct and indirect detection.

This letter is organized as follows:In the next section, we introduce the model. In section~3, we will review relic density, and  constraints which arise from  experimental observables at LEP and LHC, direct detection and indirect detection. The conclusions are given in section~4.

\section{The model}
In ITM, the matter content of SM is extended with a $SU(2)_L$ triplet scalar with $Y=0$ or $Y=2$. These additional fields are odd under $Z_2$ symmetry condition while all the SM fields own even eigenvalues.  The $Z_2$ symmetry is not spontaneously broken since the triplet does not develop a vacuum expectation value. The triplet $T$ for $Y=0$ has $\rm VEV=0$ and the SM Higgs doublet H and the triplet T scalars are defined as:
\begin{eqnarray}
T=\left(\begin{array}{cc}
   \frac{1}{\sqrt{2}}T^0 & -T^{+} \\
   -T^- & -\frac{1}{\sqrt{2}}\tz
  \end{array}\right),
\langle H\rangle=\frac{1}{\sqrt{2}} \left(\begin{array}{cc}
   0 \\
  v
  \end{array}\right),
\end{eqnarray}
where $v=246~\rm GeV$. The relevant Lagrangian which is allowed by $Z_2$ symmetry can be given by:
\begin{eqnarray}
{\cal{L}}&=&|D_{\mu}H|^2 + \tr|D_{\mu}T |^2 - V (H, T ),\nonumber\\
V (H, T )&=&m^2 |H|^2 + M^2 \tr[T^2] + \lambda_1 |H|^4\nonumber\\&+& \lambda_2 (\tr[T^2])^2+ \lambda_3 |H|^2 \tr[T^2] \label{potentioal Y=0},
\end{eqnarray}

In the case $Y=0$, ITM has three new parameters compared to the SM. We require that Higgs potential is bounded from below, which leads to following conditions on the parameters of the potential:
\begin{eqnarray}
 \lambda_1,\ \lambda_2 \geq 0 ,~~\,(\lambda_1 \lambda_2)^{1/2}-\frac{1}{2}|\lambda_3|>0\ .
\end{eqnarray}
The conditions for local minimum are satisfied if and only if $m^2 <0$, $v^2 = - m^2/2\lambda_1$ and $2M^2 + \lambda_3 v^2 > 0$. The masses of triplet scalars  can be written:
\begin{eqnarray}
m_{T^0} = m_{T^{\pm}}= \sqrt{M^2 + \frac{1}{2}\lambda_3 v^2}.\label{mDM}
\end{eqnarray}

Note that at tree level, masses of neutral and charged components are the same, but at loop level the $T^{\pm}$ are slightly heavier than $T^0$ \cite{Minimal dark matter}. The scalar and gauge interactions of ITM have been extracted in terms of real fields in \cite{ITM}. In case $Y=0$, the $Z_2$ symmetry ensures that $T^0$ cannot decay to SM fermions and can be considered as cold DM candidate. Nevertheless the Z boson can decay to $T^{\pm}$. The decay rate of $Z\rightarrow T^{\mp} T^{\pm}$ is given by:
\begin{eqnarray}
\Gamma(Z\rightarrow T^{\mp} T^{\pm}))& =\frac{g^2c^2_W m_Z}{\pi}(1-\frac{4m^2_{T^{\pm}}}{m^2_Z})^{3/2},
\label{amp1}
\end{eqnarray}
where $g$ is the weak coupling and $c_W=\cos\theta_W$. The Z boson decay width was measured by LEP experiment ($\Gamma_Z=2.4952\pm0.0023~\rm GeV$). This measurement is consistent with SM prediction. This means the Z boson decay width will strictly constrain ITM parameters space. Therefore, we assume that $m_{T^0}, m_{T^{\pm}}>45.5~\rm GeV$.

In case $Y=2$ the $SU(2)_L$ triplet can be parameterized with five new parameters\cite{ITM}. The ITM with Y=2, is already excluded by the limits from direct detection experiments. There won't be any use to study the case in this regard.

\section{ Observables and Numerical results}
In this section, we will review the relic density conditions for ITM and constraints arising from  experimental observables at LHC, direct and indirect detection.

\begin{figure}
\includegraphics[scale=0.4]{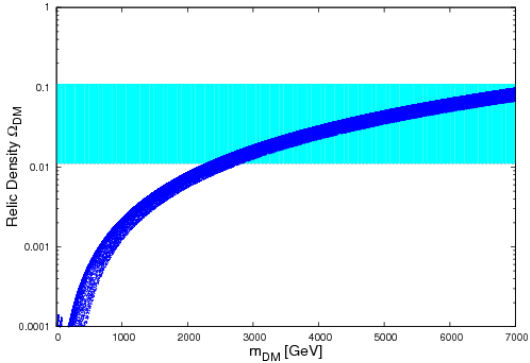} \caption{ Relic density as a function of DM mass for all the valid values of $\lambda_3$. The shadowed panel indicates regions in which $T^0$ particles contribute more than 10 percent of dark matter density.} \label{relic}
\end{figure}

\subsection{Relic Density}
\begin{figure}
\includegraphics[scale=0.4]{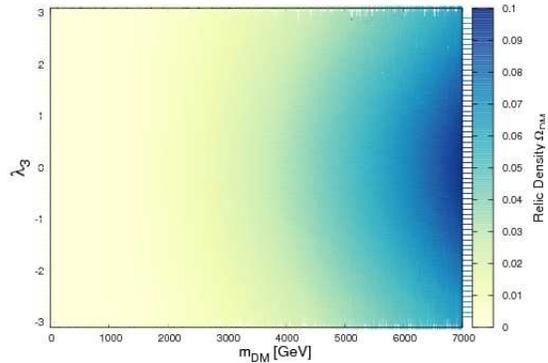} \caption{ The relic density plot in $\lambda_3$ and DM mass plane. The blue region leads to more participation in relic density.} \label{relic1}
\end{figure}

The relic density of DM is well measured by WMAP and Planck experiments and the current value is \cite{PLANCK}:
\begin{equation}
\Omega_{DM}h^2 = 0.1199 \pm 0.0027,
\label{si}
\end{equation}
where $h = 0.67 \pm 0.012$ is the scaled current Hubble parameter in units of $100\rm km/s.Mpc$. In the following, we will use this value as upper bound on the contribution of ITM in production of DM. Before the onset of freeze-out, the universe is hot and dense. As the universe expand, the temperature fall down. Ultimately $T^0$ particles will become so rare that they will not be able to find each other fast enough to maintain the equilibrium abundance. So the equilibrium ends and the freeze-out starts. Inert particles, $T^0$,  can contribute in the relic density of DM through freeze-out mechanism. Solving Boltzman equation will determine the freeze-out abundance.
We have used \textit{LanHep} \cite{lan} to generate model files which \textit{Micromega 3.2} \cite{micro} employs to calculate relic density. The relic density as a function of interaction rate changes for the different values in parameter space. Fig.~\ref{relic} and \ref{relic1} indicate how inert particles contribute in dark matter density for the different values of mass and coupling. In large mass regimes and low couplings, Inert particle can constitute whole the dark matter which is very plausible.  As it is seen in Fig.~\ref{relic} and \ref{relic1}, in context of ITM, in mass regimes lower than $7~\rm TeV$, relic density conditions are satisfied. We emphasis that for  $m_{DM}\leq2~\rm TeV$, ITM can not saturate the relic density and it demands multi-components DM to explain whole density.

\subsection{Direct Detection}
In the case of $Y=0$, DM candidate can interact with nucleon by exchanging Higgs boson. The DM-nucleon scattering cross section is given by \cite{DM lattice QCD}:
\begin{equation}
\sigma_{SI} = \frac{\lambda^2_3f_N^2}{4\pi m^4_h}\frac{m^4_N}{(m_N+m_{T^0})^2},
\label{si}
\end{equation}
where the coupling constant $f_N$ is given by nuclear matrix elements\cite{Higgs-nucleon} and $m_N=0.939~\rm GeV$ is nucleon mass.
The most strict bound on the DM-nucleon cross section obtained from $\rm LUX$\cite{LUX} experiment. The minimum upper limit on the spin independent cross section is:
\begin{eqnarray}
\sigma_{SI}&\leq& 7.6\times 10^{-46} \rm cm^2.
\label{Direct}
\end{eqnarray}

As it was mentioned in $Y=2$ case,  Due to gauge coupling of $Z$ to DM candidate, the DM-nucleon cross section is $10^{-38}~\rm cm^2$ and much larger than upper limit by LUX experiment. This excludes all the regions of parameter.

Fig.~\ref{sigmav1}, shows allowed region in DM mass and $\lambda_3$ couplings plane which does not violate $90\%$ C.L experimental upper bounds of $\rm LUX$ for $m_Z/2<m_{T^0}<m_h/2$.  In this figure, we compare these bounds with other constraints which arise from other observables.

\subsection{Invisible Higgs decays}
A SM-like Higgs boson was discovered at the LHC in 2012. Some extensions of the SM allow a Higgs particle to decay into new stable particle which is not observed by ATLAS and CMS detectors yet. For example, the Higgs boson can decay into per of  DM particles. The branching ratio of the Higgs particle to invisible particle can be used directly to  constrain parameter space of new physics. Nevertheless, invisible Higgs boson decay is not sensitive to DM coupling when $m_{T^0}>m_h/2$. In ITM, if triplet scalar mass is lighter than SM higgs boson mass, then it can contribute to the invisible decay mode of higgs boson. The total invisible Higgs boson branching ratio is given by:
\begin{eqnarray}
Br(h\rightarrow \rm Invisible)& =\frac{\Gamma(h\rightarrow \rm Inv)_{\rm SM}+\Gamma(h\rightarrow 2T^0)}{\Gamma(h)_{\rm ITM}},
\label{decayinv}
\end{eqnarray}
where $\Gamma(h)_{\rm SM}=4.15 ~ \rm MeV$ \cite{SM Higgs branching ratio} is total width of higgs boson in SM and $\Gamma(h)_{\rm ITM}$ is total decay width of higgs boson in ITM:
\begin{eqnarray}
\Gamma(h)_{\rm ITM}=\Gamma(h)_{\rm SM}+\sum_{\chi=T^0,T^{\pm},\gamma}\Gamma(h\rightarrow 2\chi).
\label{Total}
\end{eqnarray}
The partial width for $h\rightarrow 2T^0$ and $h\rightarrow T^{\pm}T^{\pm}$ are given by:
\begin{eqnarray}
\Gamma(h\rightarrow 2T^0)& =\frac{\lambda^2_3v^2_0}{4\pi m_h}\sqrt{1-\frac{4m^2_{T^0}}{m^2_h}},\nonumber\\
\Gamma(h\rightarrow T^{\pm}T^{\pm})& =\frac{\lambda^2_3v^2_0}{\pi m_h}\sqrt{1-\frac{4m^2_{T^{\pm}}}{m^2_h}},
\label{decayinv1}
\end{eqnarray}
and $h\rightarrow 2\gamma$ was given in \cite{Ayazi:2014tha}. The SM branching ratio for the decay of Higgs to invisible particles is  $1.2\times10^{-3}$ which is produced by $h\rightarrow ZZ^*\rightarrow 4\nu$  \cite{Higgs branching ratio}. A search for evidence of invisible decay mode of a Higgs boson has done by ATLAS collaboration and an upper limit of $75\%$ with $95\%$ C.L is set on branching ratio of Higgs boson invisible mode \cite{Invisible Higgs decay Exp}. Since invisible higgs decay is forbidden kinematically for $m_{D}>m_{h/2}$, we present our results for $Br(h\rightarrow \rm Invisible)$ and other observables only for $m_Z/2<m_{T^0}<m_h/2$. In Fig.~\ref{sigmav1}, we suppose $m_Z/2<m_{T^0}<m_h/2$ and depict allowed region in mass of DM and $\lambda_3$ coupling plane which is consistent with  experimental upper limit on $Br(h\rightarrow \rm Invisible)$(with $95\%$ C.L). It is remarkable that  valid area of $Br(h\rightarrow \rm Invisible)$ and direct detection experiments are very similar.

\subsection{Annihilation of Dark Matter into monochromatic gamma-ray}

\begin{figure}
\includegraphics[scale=0.45]{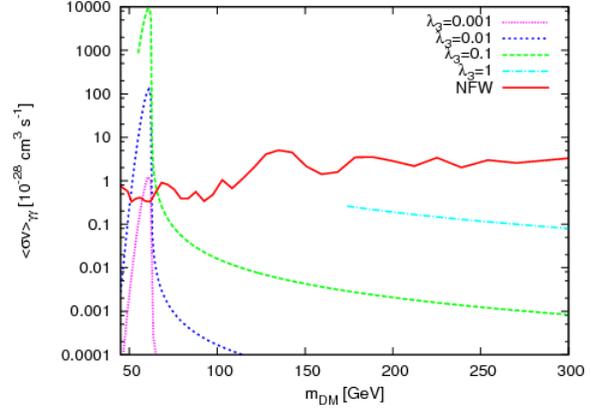} \caption{The thermal average annihilation cross-section of $T^0$ (DM) to $\gamma\gamma$ as a function of
the DM mass for several values of $\lambda_3$. The solid red lines shows the upper limits on annihilation cross-section which have borrowed from \cite{FermiLat}.} \label{sigmav}
\end{figure}

DM particles annihilation or decay can produce monochromatic photon and contribute to the diffuse gamma-ray background. In ITM, $T^{\pm}$ can contribute to annihilation of DM candidate into monochromatic photons $2T^0\rightarrow 2\gamma$. The amplitude of possible annihilation of DM candidate in ITM into $2\gamma$ has been calculated in \cite{Ayazi:2014tha}.

\begin{figure}
\includegraphics[scale=0.38]{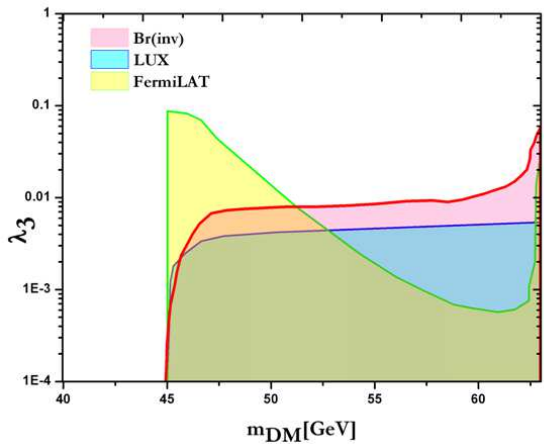} \caption{ Shaded areas depict ranges of parameter space in mass of DM and $\lambda_3$ coupling plane which are consistent with
 experimental measurements of $Br(h\rightarrow \rm Invisible)$, upper limit on $\sigma_{\rm FermiLAT}$ (indirect detection) and $\sigma_{\rm LUX}$ (direct detection).} \label{sigmav1}
\end{figure}

Flux upper limits for diffuse gamma-ray background and  gamma-ray spectral lines from $7$ to $300~\rm GeV$  obtained from 3.7 years data has been presented by FermiLAT collaboration in \cite{FermiLat}. In this section, we obtain thermal average cross section of annihilation and apply these data to set constrain on ITM parameter space. In Fig.~\ref{sigmav}, we display  the thermal average cross section for annihilation of DM to $\gamma\gamma$ as a function of the DM mass for several values of $\lambda_3$. For process $2T^0\rightarrow \gamma\gamma$, we assume $E_{\gamma}=m_{DM}$. The solid red lines depicts the upper limits on annihilation cross-section for NFW density profile in the Milky Way which have borrowed from \cite{FermiLat}. In this figure, for $m_{DM}>63~\rm GeV$, total annihilation cross-section is much lower than FermiLAT upper limit. This means  FermiLAT data can not constrain ITM parameters space in this region. Nevertheless, for low DM mass ($m_{DM}<63~\rm GeV$ near to the pole of Higgs propagator at  $m_{DM}=m_h/2$), the annihilation cross section increases and will be larger than upper limit. To study this phenomena, we consider the minimum upper limit on $\sigma_{\rm FermiLAT}=0.33\times 10^{-28}$ for NFW profile \cite{FermiLat} in Fig.~\ref{sigmav1} and depict allowed regions on DM mass and $\lambda_3$ coupling plane which are consistent with this limit. We compared all results for direct search, invisible Higgs decay and indirect search in this figure. It is remarkable that indirect search constraint is stronger than direct detection limit in region $52<m_{DM}<63$.

\section{Concluding remarks}
In this talk, we have presented an extension of SM which includes a $SU(2)_L$ triplet scalar with hypercharge $Y=0,2$. This model provide suitable candidate for DM, because the lightest component of triplet field is neutral and for the $m_{DM}<7~\rm TeV$, conditions of relic abundance are satisfied. We focus on  parameter space  which is allowed by PLANCK data and study collider phenomenology of inert triplet scalar DM at the LHC.

We have shown that the effect of ITM on invisible Higgs decay for low mass DM ($m_{DM}<63~\rm GeV$) can be as large as constraints from LUX direct detection  experiment ( see Fig.~\ref{sigmav}-b).

We consider the annihilation cross section of DM candidate into $2\gamma$. The minimum upper limit on annihilation cross-section from FermiLAT have been employed to constraint parameters space of ITM. We also compared our results with constraints from direct detection and showed for $52<m_{DM}<63~\rm GeV$,  FermiLAT constraint is stronger than direct detection constraint for low mass DM.

\textbf{\ \ \ \  \ \ \ \ \ \ \ \ \ \  Acknowledgement}

We would like to thank the organizer of "From Higgs to Dark Matter" conference held at Geilo, Norway (14-17 December 2014) where this talk
was presented.

\end{document}